\documentclass[%
 reprint,
 amsmath,amssymb,
 aps,
]{revtex4-1}

\usepackage[dvipdfmx]{graphicx}
\usepackage{dcolumn}
\usepackage{bm}
\usepackage{color}

\begin{document}

\preprint{APS/123-QED}

\title{Decomposition of scattering phase shifts and reaction cross sections using the complex scaling method}

\author{Myagmarjav ODSUREN}
\email {odsuren@nucl.sci.hokudai.ac.jp}
\affiliation{Meme Media Laboratory, Hokkaido University, Sapporo 060-8628, Japan}
\affiliation{Nuclear Research Center, National University of Mongolia, Ulaanbaatar 210646, Mongolia}

\author{Kiyoshi KAT\=O}
\email{kato@nucl.sci.hokudai.ac.jp}
\affiliation{Faculty of Science, Hokkaido University, Sapporo 060-0810, Japan}

\author{Masayuki AIKAWA}
\email{aikawa@sci.hokudai.ac.jp}
\affiliation{Faculty of Science, Hokkaido University, Sapporo 060-0810, Japan}

\author{Takayuki MYO}
\email{myo@ge.oit.ac.jp}
\affiliation {General Education, Faculty of Engineering, Osaka Institute of Technology, Osaka 535-8585, Japan}
\affiliation {Research Centre for Nuclear Physics (RCNP), Osaka University, Ibaraki 567-0047, Japan}

\date{\today}

\begin{abstract}
We apply the complex scaling method to the calculation of scattering phase shifts and extract the contributions of resonances in a phase shift and a cross section. The decomposition of the phase shift is shown to be useful to understand the roles of resonant and non-resonant continuum states. As examples, we apply this method to several two-body systems: (i) a schematic model with the Gyarmati potential which produces many resonances, (ii) the $\alpha-\alpha$ system which has a Coulomb barrier potential in addition to an attractive nuclear interaction, and (iii) the $\alpha-n$ system which has no barrier potential. Using different kinds of potentials, we discuss the reliability of this method to investigate the resonance structure in the phase shifts and cross sections.
\begin{description}
\item[PACS numbers] 21.60.Gx, 24.10.-i
\end{description}
\end{abstract}

\pacs{Valid PACS appear here}
\maketitle


\section{Introduction}
Nuclear scattering is the most important phenomena from which we can obtain information and knowledge on various nuclear properties. Many theoretical approaches and experimental techniques have been developed to extract physics from the scattering phenomena. Resonances observed as some peaks in scattering cross sections provide us with the precious information for understanding nuclear interactions and structures. In particular, it is indispensable in recent developments of unstable nuclear physics to investigate the resonances involved in the unbound states locating above the many-particle decay thresholds, because the unstable nuclei barely have bound states and most of the excited states are resonances. Furthermore, to understand weakly bound states in unstable nuclei, such as a halo structure, we need to investigate continuum states together with bound states because of a strong coupling between them \cite{Ho12}.

Study of resonances in the scattering problem of light nuclei has been carried out by using various methods, one of which is the complex scaling method (CSM) \cite{Ag71}. The theory of the complex scaling was proposed mathematically \cite{Ag71} and it has been extensively applied to the atomic and nuclear physics \cite{Ho83,Mo98,Ao06}. In the CSM, resonant states of the many-body systems are described using the appropriate $L^2$ basis functions. The resonance wave functions are obtained as eigenstates together with bound states by carrying out the diagonalization of the complex scaled Hamiltonian. Hence, we can calculate expectation values of physical quantities for resonances \cite{Ho97}. Recently, resonances decaying up to five-body systems have been studied in the CSM and successfully compared with experiments \cite{My11,My12}.

We have also discussed that the CSM is very useful in studies of not only resonant states but also scattering states \cite{Ao06}. The complex scaling separates the resonant states with resonance energies $E^{res}$ and widths $\Gamma$ ($\tan^{-1}(\Gamma/2E^{res})<2\theta$) from the continuum states obtained on a '$2\theta$-line', where $\theta$ is a scaling angle in the CSM \cite{Ag71}. We can extract scattering properties from the continuum solutions on the '$2\theta$-line' together with the resonance solutions. Calculations of the phase shifts of continuum states with complex energies along the '$2\theta$-line' provide the background phase shifts exclusive of the resonances existing in the energy region between the '$2\theta$-line' and the real energy axis \cite{Ho97}. As examples of applications of the CSM to scattering problems, response functions for external electric fields and breakup cross sections of two-neutron halo nuclei have been discussed and shown for observed data to be well described by a two-neutron-plus-core three-body model \cite{My98,My07,My01}. In their results, it is concluded that a sequential breakup process is dominant rather than a direct three-body breakup.

In these calculations, a complex scaled Green's function has been introduced to obtain the response functions as observables. We can also calculate scattering quantities such as phase shifts in a form of sum of resonances and background terms. It is possible to investigate the resonance contributions and to obtain a deep understanding of resonance structure by separation of a scattering quantity. Suzuki {\it et al.} \cite{Su05} showed that scattering phase shifts can be calculated from the continuum level density (CLD) which is expressed using the complex scaled Green's function. The CLD is also called a time delay and has a relation with the $S$-matrix \cite{Le69,Sh92}. In the CSM, we obtain the discretized continuum states as complex energy eigenstates. Using those complex eigen-energies, we can evaluate the CLD as a smoothed real-energy function.

In this paper, we discuss the explicit relation between the scattering phase shifts and complex-energy eigenvalues in the CSM via the CLD. The results provide us with deeper understanding of the role of resonant states characterised by the widths described as an imaginary part of the eigen-energy. We show the results of typical potential scattering which has many resonances near the real energy. We also analyze the several realistic systems and compare the results with the observed data. The observed scattering cross sections have various energy distributions as a result of interference between resonance and background terms of phase shifts \cite{Fa61}. We see this interference in the $\alpha-n$ and $\alpha-\alpha$ systems.

The outline of the paper is as follows. In Section II, the method is briefly explained, and the scattering phase shifts are explicitly shown by using the complex energy solutions in the CSM. In Section III, the present method is applied to several two-body systems; a typical potential problem, so called Gyarmati potential \cite{Gy90} is discussed. The phase shifts of $\alpha-\alpha$ and $\alpha-n$ systems are presented together with experimental ones. Finally, in Section IV, discussion and summary are given.

\section{Decomposition of phase shifts}
\subsection{Complex scaling method}
We take up two-body systems which are described by the Schr\"odinger equation
\begin{equation}
H\Psi=E\Psi,
\end{equation}
where the Hamiltonian $H$ consists of kinetic energy $T$ and potential $V$ for the relative motion between two bodies. The eigenvalue problem is generally solved under a boundary condition of asymptotic out-going waves for bound states and resonances. The out-going boundary condition directly enables us to solve bound states in an $L^2$ functional basis set because the states have negative energies and a damping behavior in the asymptotic region. Resonant states are unbound and defined as the eigenstates belonging to the complex eigen-energy which corresponds to a complex momentum value in the lower half plane (unphysical plane \cite{Ao06}). The resonant states cannot be solved in the $L^2$ functional space due to asymptotically divergent behavior. Furthermore, continuum states of arbitrary positive energies cannot also be obtained under the out-going condition.

The complex scaling has been introduced to solve resonant states within $L^2$ basis functions, and is defined by the following complex-dilatation transformation for relative coordinate $\vec{r}$ and momentum $\vec{k}$ \cite{Ag71};
\begin{equation}
\vec{r}\to\vec{r}e^{i\theta},\hspace{2cm}\vec{k}\to\vec{k}e^{-i\theta},
\end{equation}
where $\theta$ is a scaling angle and $0<\theta < \theta_{max}$. The maximum value $\theta_{max}$ is determined so as to keep analyticity of the potential. For example, $\theta_{max}=\pi/4$ for a Gaussian potential. This transformation makes every branch cut rotated by $-2\theta$ on the complex energy plane. In the wedge region pinched by the rotated branch cut and the positive energy axis, resonance eigenstates are obtained by solving the following eigenvalue problem
\begin{eqnarray}
\sum_{j=1}^{N}\left\langle \phi_i|H(\theta)|\phi_j\right\rangle c_j^\alpha(\theta)&=& E_\alpha\sum_{j=1}^{N}\left\langle \phi_i|\phi_j\right\rangle c_j^\alpha(\theta),\nonumber \\
\Psi^\alpha(\theta)&=&\sum_{i=1}^Nc_i^\alpha(\theta)\phi_i,
 \label{eq2-3}
\end{eqnarray}
within an appropriate non-orthogonal $L^2$ basis set $\{\phi_i, i=1,2,\cdots,N\}$. The index $\alpha$ is to distinguish the eigenstates $\Psi^\alpha(\theta)$ of the complex scaled Hamiltonian $H(\theta)$. The bound states are obtained on the negative energy axis independently from $\theta$ as well as the ordinary bound states. Because of a finite number of basis states, the continuum states are discretized with complex energies distributed on the rotated branch cut ($2\theta$-line).

The eigenvalues and eigenstates of the complex scaled Schr\"odinger equation (\ref{eq2-3}) are classified as
\begin{widetext}
\begin{eqnarray}
(E_\alpha, \Psi^\alpha(\theta))=\left\{\begin{array}{cll}
                                        (E_b,\ \Psi^b)  & b=1,\cdots,N_b  & \mbox{; bound states}     \\
                                        (E_r,\ \Psi^r)  & r=1,\cdots,N_r^{\theta} & \mbox{; resonant states} \\
                                        (E_c(\theta),\ \Psi^c) & c=1,\cdots,N-N_b-N_r^{\theta}  &\mbox{; continuum states}
                               \end{array}, \right.
\end{eqnarray}
\end{widetext}
where $N_b$ and $N_r^{\theta}$ are the number of bound states and the number of resonant states which depends on $\theta$, respectively. The complex energies of resonant states are obtained as $E_r=E_r^{res}-i\Gamma_r/2$, when $\tan^{-1}{(\Gamma_r/2E^{res}_r)} < 2\theta$. The discretized energies $E_c(\theta)$ of continuum states are $\theta$-dependent and expressed as $E_c(\theta)=\epsilon_c^r-i\epsilon_c^i$.

These three-kinds solutions of the complex-scaled Schr\"odinger equation construct the extended completeness relation \cite{My98};
 \begin{eqnarray}
\sum_{b=1}^{N_b}|\Psi^b \rangle\langle\tilde{\Psi}^b|&+& \sum_{r=1}^{N_r^{\theta}}|\Psi^r\rangle\langle\tilde{\Psi}^r| \nonumber \\
                                                     &+& \int_{L_c}dE_c |\Psi^c\rangle\langle\tilde{\Psi}^c|=1,  \nonumber \\
\end{eqnarray}
where the tilde ( $\tilde{ }$ ) in bra-states means the bi-orthogonal states with respect to the ket-states due to non-hermitian property of $H(\theta)$. The integration of the third term is taken along the rotated branch cut $L_c$. The extended completeness relation has been proven for single- and coupled-channel systems \cite{Gi03,Gi04}. In the case of eigenstates within a finite number of $L^2$ basis states, the integration for continuum states is approximated by the summation of discretized states as \cite{My98}
 \begin{eqnarray}
\sum_{b=1}^{N_b}|\Psi^b \rangle\langle\tilde{\Psi}^b|&+& \sum_{r=1}^{N_r^{\theta}}|\Psi^r\rangle\langle\tilde{\Psi}^r| \nonumber \\
                                                     &+& \sum_{c=1}^{N-N_b-N_r^{\theta}} |\Psi^c\rangle\langle\tilde{\Psi}^c|\approx 1.
                                                      \label{eq2-4}
\end{eqnarray}
It has been investigated that the reliability of the approximation of the continuum states are confirmed by using a sufficient large basis number of $N$ in the CSM \cite{Ao06, Su05}.

\subsection{Continuum level density and scattering phase shift}
The CLD, $\Delta (E)$ as function of the real energy $E$ is defined as \cite{Le69,Sh92}
\begin{equation}
\Delta (E)=-\frac{1}{\pi}\text{Im}\{\text{Tr}[G^+(E)-G_0^+(E)]\},\label{eq2-2-1}
\end{equation}
where
\begin{equation}
G^+(E)=(E+i\epsilon-H)^{-1}, \hspace{0.3cm} G_0^+(E)=(E+i\epsilon-H_0)^{-1} \nonumber
\end{equation}
are the full and free Green's functions, respectively.
The CLD is also related to the scattering phase shift $\delta(E)$ and the relation is expressed in a single channel case as \cite{Le69,Sh92}:
\begin{equation}
\Delta(E)=\frac{1}{\pi}\frac{d\delta(E)}{dE}.\label{eq2-2-2}
\end{equation}

Using  Eqs. (\ref{eq2-2-1}) and (\ref{eq2-2-2}), we can obtain the phase shift $\delta(E)$ in terms of the eigenvalues of $H$ and $H_0$ by integrating the CLD over the energy $E$. When we apply the complex scaling and obtain the complex scaled Green's function, the CLD can be expressed as
 \begin{eqnarray}
\Delta (E)&=&-\frac{1}{\pi}\text{Im}\{\text{Tr}[\frac{1}{E-H(\theta)}-\frac{1}{E-H_0(\theta)}]\}.   \label{eq2-2-3}
\end{eqnarray}
Furthermore, we apply the extended completeness relation given in Eq. (\ref{eq2-4}) to the calculation of $\Delta(E)$ in Eq. (\ref{eq2-2-3}), where we use the solutions obtained by diagonalization of matrix elements of $H(\theta)$ and $H_0(\theta)$ with a finite number $N$ of basis functions. The energy eigenvalues of $H_0(\theta)$ are given as $\epsilon_k^{0r}-i\epsilon_k^{0i}$ with $k=1, \cdots,N$. All of these values are distributed on the '$2\theta$-line'. The CLD of Eq. (\ref{eq2-2-3}) is approximated as
\begin{eqnarray}
\Delta(E)&\approx& \Delta^N_\theta(E) \nonumber \\
&=& -\frac{1}{\pi}\text{Im}\left[\sum_{b=1}^{N_b}\frac{1}{E+i0-E_b}\right.\nonumber \\
& & \hspace{0.5cm} +\sum_{r=1}^{N_r^\theta}\frac{1}{E-E_r^{res}+i\Gamma_r/2} \nonumber \\
& & \hspace{1.3cm} +\sum_{c=1}^{N_c^\theta}\frac{1}{E-\epsilon^r_c+i\epsilon^i_c} \nonumber \\
& & \hspace {2cm} \left.  -\sum_{k=1}^{N}\frac{1}{E-\epsilon^{0r}_k+i\epsilon^{0i}_k}\right], \nonumber \\
& &
   \label{eq2-2-4}
\end{eqnarray}
where $N=N_b+N_r^\theta+N_c^\theta$. It is important to note that the approximated CLD, $\Delta^N_\theta(E)$ basically has a dependence on the scaling angle $\theta$ because we employ a finite number $N$ of the basis states to expand the complex scaled wave functions. In the calculation we adopt a sufficiently large number of $N$ to keep the numerical accuracy and to make the $\theta$-dependence negligible in the solutions \cite{Su05}. Thus we calculate the phase shift from $\Delta^N_\theta(E)$;
\begin{eqnarray}
\delta^N_\theta(E)&=&\pi\int_{-\infty}^{E} \Delta^N_\theta(E)dE  \nonumber \\
&=&\int_{-\infty}^{E}dE\ \left[\sum_{b=1}^{N_b}\pi\delta(E-E_b) \right.\nonumber \\
& & \hspace{0.5cm} +\sum_{r=1}^{N_r^\theta}\frac{\Gamma_r/2}{(E-E_r^{res})^2+\Gamma_r^2/4} \nonumber \\
& & \hspace{1.3cm} +\sum_{c=1}^{N_c^\theta}\frac{\epsilon^i_c}{(E-\epsilon^r_c)^2+(\epsilon^{i}_c)^2} \nonumber \\
& & \hspace{2cm}\left.  -\sum_{k=1}^{N}\frac{\epsilon^{0i}_k}{(E-\epsilon^{0r}_k)^2+(\epsilon^{0i}_k)^2}\right]. \nonumber \\
\label{eq2-2-5}
\end{eqnarray}

\begin{figure}[!htb]
\centering
\noindent
\includegraphics[width=0.9\columnwidth, height=0.5\textheight]{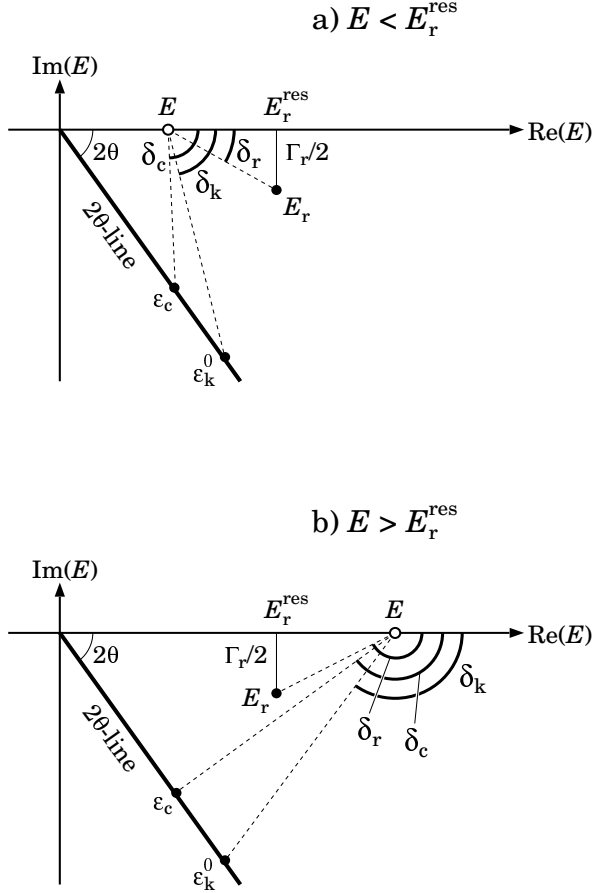}
\caption{The geometrical expressions for phase shifts: $\delta_r$, $\delta_c$ and $\delta_k$ as functions of the energy $E$. Both $E < E^{res}_r$ and $E > E^{res}_r$ cases are displayed in the upper panel a) and lower panel b), respectively. The details are explained in the text.}
\label{fig00}
\end{figure}

Performing integration of every term, we obtain the following expression:
\begin{eqnarray}
\delta^N_\theta(E)&=&N_b\pi+\sum_{r=1}^{N_r^\theta}\left\{-\cot^{-1}\left(\frac{E-E_r^{res}}{\Gamma_r/2}\right)\right\} \nonumber \\
& & \hspace{1 cm}+\sum_{c=1}^{N_c^\theta}\left\{-\cot^{-1}\left(\frac{E-\epsilon^r_c}{\epsilon^{i}_c}\right)\right\} \nonumber \\
& &\hspace{2 cm} -\sum_{k=1}^{N}\left\{-\cot^{-1}\left(\frac{E-\epsilon^{0r}_k}{\epsilon^{0i}_k}\right)\right\}, \nonumber \\
& &
 \label{eq2-2-6}
\end{eqnarray}
where we assume $E\ge 0$. When we define $\delta_r$, $\delta_c$ and $\delta_k$ as
\begin{eqnarray}
\centering
\cot{\delta_r}&=&\frac{E^{res}_{r}-E}{\Gamma_r/2},  \nonumber \\
\cot{\delta_c}&=&\frac{\epsilon_c^r-E}{\epsilon_c^i}, \nonumber \\
\cot{\delta_k}&=&\frac{\epsilon_k^{0r}-E}{\epsilon_k^{0i}},  \label{eq2-2-7}
\end{eqnarray}
respectively, we express the phase shift as
\begin{equation}
\delta^N_\theta(E)=N_b\pi+\sum_{r=1}^{N^\theta_r}\delta_r+\sum_{c=1}^{N^\theta_c}\delta_c-\sum_{k=1}^{N}\delta_k.
\label{eq2-2-8}
\end{equation}
The geometrical expressions of $\delta_r$, $\delta_c$ and $\delta_k$ are given for $E < E^{res}_r$ or $E > E^{res}_r$ in Fig.~\ref{fig00}.
The resonance phase shift $\delta_r$ is the angle of the $r$th resonant pole at an energy $E$ on the real energy axis. At the resonance energy, $E=E^{res}_r$, the relation $\delta_r=\pi/2$ is confirmed for every resonant pole. In addition, $\delta_r=\tan^{-1}(\Gamma_r/2E_r^{res})>0$ at $E=0$ and $\delta_r=\pi$ at $E=\infty$ for each resonance. Similarly the phase shifts from continuum terms including asymptotic part, $\delta_k$ are given by the angles of the discretized continuum energies. At $E=\infty$, the continuum terms of the phase shifts go to $-(N_b+N^\theta_r)\pi$ because of the relation $N=N_b+N_r^{\theta}+N_c^{\theta}$. Thus, $\delta^N_\theta\to 0$ for $E\to \infty$ and the Levinson theorem is confirmed as
\begin{equation}
\delta^N_\theta(E=0)-\delta^N_\theta(E\to \infty)=N_b \pi,
\end{equation}
where the number of the Pauli forbidden states is included in $N_b$ when they exist.

The cross section is described by using these phase shifts, and we can see the contributions from both resonant poles and continuum terms. When we concentrate our interest on the contribution from a single resonant pole, other terms are described as a background phase shift. We can have the similar discussion as was done by Fano \cite{Fa61} because the partial cross section $\sigma_\ell(E)$ for the orbital angular momentum $\ell$ is given as
\begin{equation}
\sigma_\ell(E)=\frac{4\pi(2\ell+1)}{k^2}\sin^2\delta_\ell(E),\label{eq2-2-9}
\end{equation}
where $k^2=2E\mu/\hbar^2$ with the reduced mass $\mu$. The phase shift $\delta_\ell(E)$ is expressed in the form as $\delta_r+\delta_B$, where $\delta_r$ and $\delta_B$ are the single resonance and the background terms including all other terms given in Eq. (\ref{eq2-2-8}), respectively. The shape of the cross section can be investigated by evaluating the resonance ($\delta_r$) and background ($\delta_B$) phase shifts.

\section{Applications to several systems}
\subsection{Typical potential scattering}
 We apply the method of analyzing the phase shifts to a simple schematic potential which is introduced in Ref. ~\cite{Gy90}. The explicit form of the Hamiltonian is given as
\begin{equation}
H=-\frac{\hbar^2}{2\mu}\nabla^2+V(r),\label{eq3-1-1}
\end{equation}
where the so-called Gyarmati potential is
\begin{equation}
V(r)=-8.0 \exp(-0.16r^{2})+4.0 \exp(-0.04r^{2}).
\label{eq3-1-2}
\end{equation}
It is assumed that $\hbar^2/ \mu=1$ (MeV$\cdot \text{fm}^2$).

This potential has an attractive pocket in the inside and a repulsive barrier in the outside. In this system, one bound state and several resonant poles for $J^\pi=0^+$ and $1^-$ are obtained \cite{Ho97,My97}. It is interesting to see the contributions from those resonant poles to the scattering quantities. In Ref. ~\cite{Ho97}, it has been shown that $E1$-transition strengths into the resonant states exhaust the sum rule value. In Ref.~\cite{My97}, continuum states on the '$2\theta$-line' of the complex energies have been investigated through the scattering phase shifts on the rotated branch cut. The calculated phase shifts for different $\theta$ values suggest that the resonant states located above the $2\theta$-line behave like bound states. The phase shifts approach to $-n\pi$ in the higher energy region where $n$ is the number of those resonant states as will be discussed in Section IV.

To solve the eigenvalue problem of Eq.~(\ref{eq2-3}), we employ the Gaussian basis functions \cite{Hi03} given as
\begin{equation}
\phi_{i}(r)=N_{\ell}(b_{i})r^{\ell}\exp \left(-\frac{1}{2b^{2}_{i}}r^{2}\right) Y_{\ell m}(\hat{r}),
\label{eq3-1-3}
\end{equation}
where the range parameters are given by a geometric progression as $b_i=b_0\gamma^{i-1};\ i=1\cdots,N$, and the normalization factor $N_{\ell}(b_i)$ is given as
\begin{equation}
N_{\ell}(b_i)=\sqrt{\frac{2^{\ell+2}}{\sqrt{\pi}b_i^{2\ell+3}(2\ell+1)!!}}.
\label{eq3-1-4}
\end{equation}
We take $N=20$ and employ the optimal values of $b_0$ and $\gamma$ so as to obtain stationary resonance solutions. The results of eigenvalues for bound and resonant states obtained with $\theta=20^\circ$ are shown in Table~I. The results are same as those in Refs.~\cite{Ho97,My98}. In addition, the consistent and stable results were obtained for the CLD at different scaling angles ($\theta=10^\circ, 15^\circ$ and $20^\circ$) in Fig. 4 of Ref.~\cite{Su05}. Using those eigenvalues including the continuum states, we calculate phase shifts by Eq.~(\ref{eq2-2-8}).
\begin{table}[th]
 \centering
\caption{Bound and resonance energies with decay widths calculated with $\theta=20^\circ$ for the $J^{\pi}=0^{+}$ and $1^{-}$ partial waves.}
\label{tab1}
\begin{tabular}{ll|ll}
\hline
\multicolumn{2}{c|}{$0^{+}$ wave} & \multicolumn{2}{c}{$1^{-}$ wave} \\
\hline
\hline
E (MeV) & state & E (MeV) & state  \\
\hline
-1.928 & bound  & -0.675 & bound \\
0.310-$i10^{-6}$  & resonance & 1.171-$i0.005$  & resonance\\
1.632-$i$0.123 &  resonance  & 2.031-$i0.489$  & resonance\\
2.249-$i$1.040 &  resonance  & 2.832-$i1.199$  & resonance\\
2.854-$i$2.570 & resonance  & 3.934-$i1.788$  & resonance \\
\hline
\end{tabular}
\end{table}

\begin{figure}[bh]
\centering
\includegraphics[width=1.01\columnwidth, height=0.3\textheight]{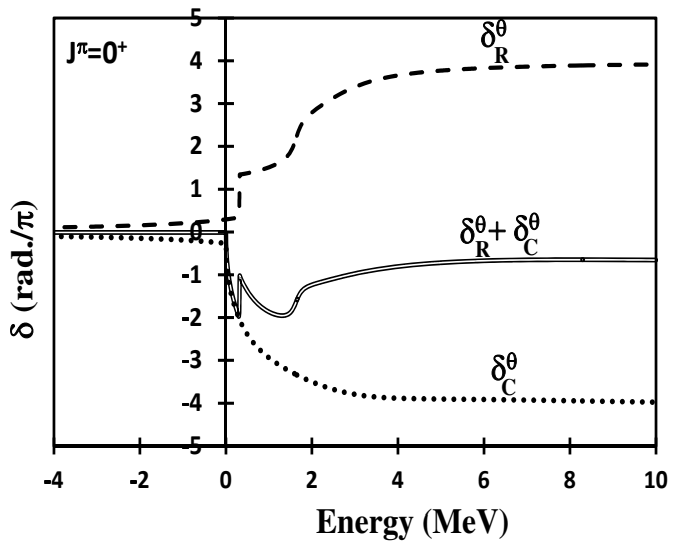}
\caption{The resonance $(\delta^\theta_R)$ and continuum $(\delta^\theta_C)$ phase shifts and the sum $\delta^\theta_R+\delta^\theta_C$ for $\theta=20^\circ$.}
\label{fig3-1-1}
\end{figure}
\begin{figure}[th]
\centering
\includegraphics[width=1.02\columnwidth, height=0.3\textheight]{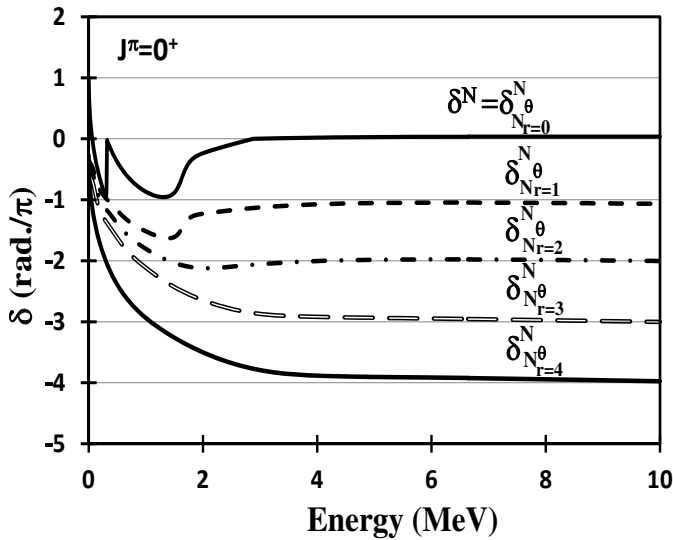}
\caption{The phase shifts of the Gyarmati potential for $J^{\pi}=0^{+}$ and the subtraction of the resonance terms one by one.}
\label{fig3-1-2}
\end{figure}

\begin{figure*}[!tb]
\centering
\hspace*{-4.2cm}
\includegraphics[width=0.7\textwidth,height=0.50\textwidth]{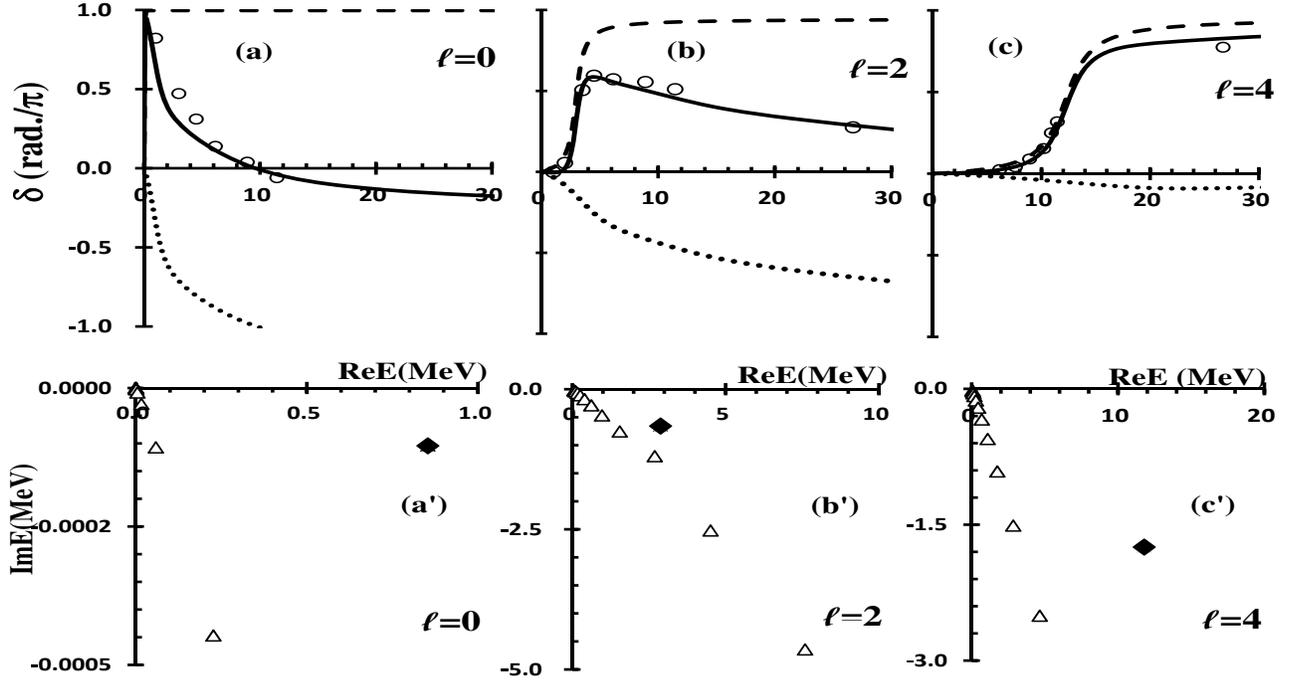}
\caption{Upper panel: The decomposition of scattering phase shifts of the $\alpha-\alpha$ system for (a) $\ell=0$, (b) $2$ and (c) $4$, respectively. The dashed-lines and dotted-lines represent the contributions of the resonance and continuum terms, respectively. The solid-lines display total scattering phase shifts. The experimental data \cite{He10} are shown with open circles. Lower panel: The distributions of eigenvalues are given in the complex energy plane for each partial wave.}
\label{fig3-2-1}
\end{figure*}
\begin{figure*}[tb]
\includegraphics[width=0.80\textwidth,height=0.5\textheight]{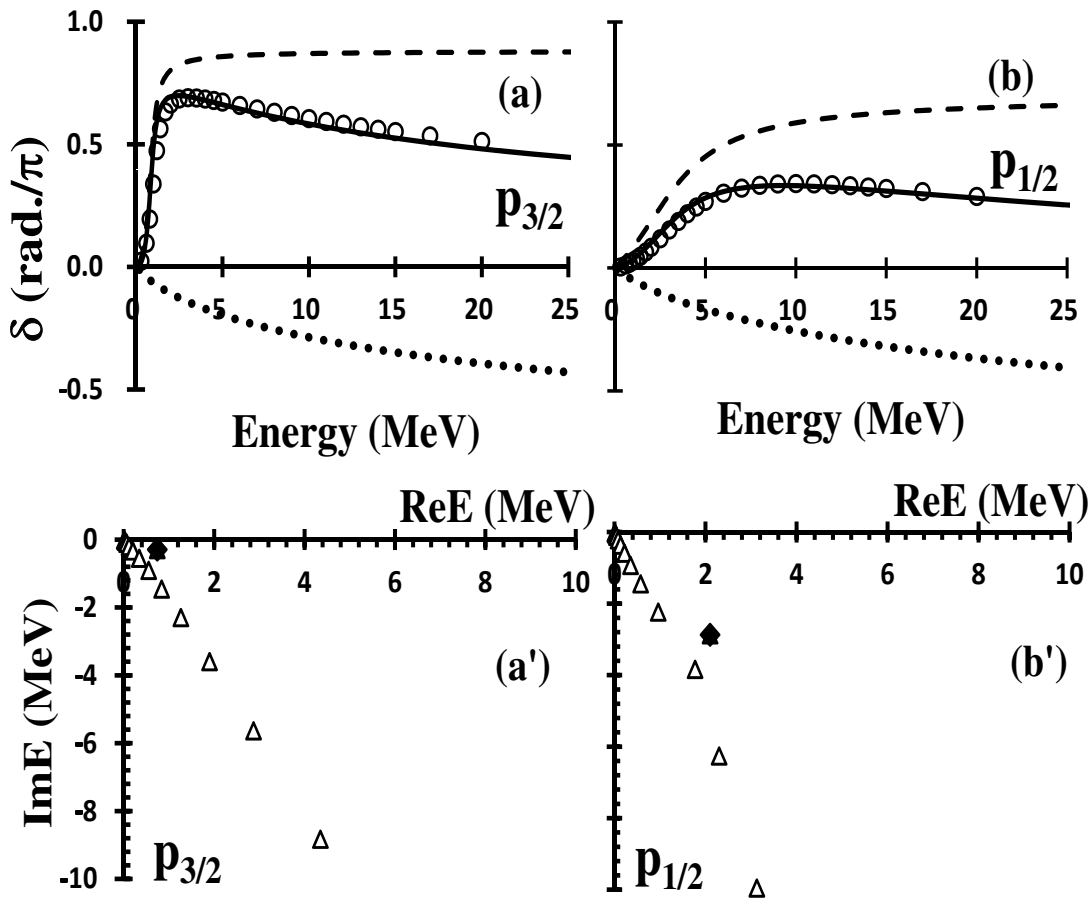}
\caption{Upper panel: The decomposition of the scattering phase shifts of $\alpha-n$ system for (a) $p_{3/2}$ and (b) $p_{1/2}$. The dashed-lines and dotted-lines represent the contributions of the resonance and continuum terms, respectively. The solid-lines display total scattering phase shifts. The experimental data \cite{Ho66} are shown with open circles. Lower panel: The distributions of eigenvalues are displayed in the complex energy plane.}
\label{fig3-2-2}
\end{figure*}
\begin{figure}[h]
\includegraphics[width=0.99\columnwidth, height=0.31\textheight]{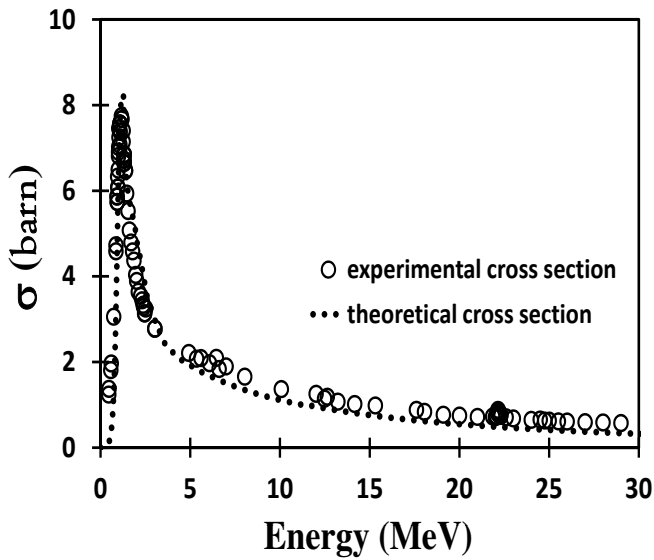}
\caption{Total cross sections as functions of relative energy of the $\alpha-n$ system. The dotted-line shows the present calculation and open circles are experimental data ~\cite{Au62}.}
\label{fig3-3-1}
\end{figure}
\begin{figure*}[!tb]
\includegraphics[width=0.80\textwidth,height=0.35\textheight]{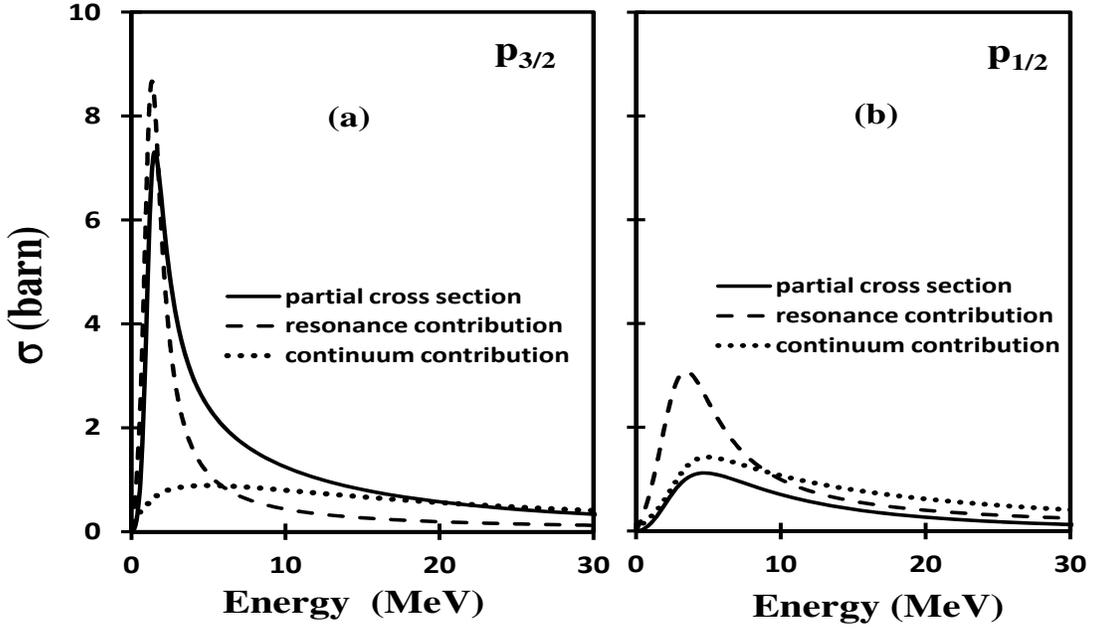}
\caption{The partial cross sections of the (a) $p_{1/2}$ and (b) $p_{3/2}$ waves. The dashed, dotted and solid-lines denote the results of resonance, continuum contributions and partial cross sections, respectively.}
\label{fig3-3-2}
\end{figure*}
In Figs.~\ref{fig3-1-1} and \ref{fig3-1-2}, we show the calculated phase shifts of the $J^\pi=0^+$ state. Here, we define resonance and continuum phase shifts, $\delta^{\theta}_R$ and $\delta^{\theta}_C$ as
\begin{equation}
\delta^\theta_R=\sum_{r=1}^{N_r^\theta}\delta_r,\hspace{1cm}\delta^\theta_C=\sum_{c=1}^{N_c^\theta}\delta_c-\sum_{k=1}^N\delta_k,
\label{eq3-1-5}
\end{equation}
respectively, where $\delta_r$, $\delta_c$ and $\delta_k$ are given in Eq. (\ref{eq2-2-7}). The phase shifts, $\delta^\theta_R$ and $\delta^\theta_C$ have the $\theta$-dependence coming from the number $N^\theta_r$ of resonant poles above the '$2\theta$-line'. Both of two-kinds of phase shifts $\delta^\theta_R$ and $\delta^\theta_C$ do not show zero in the negative energy region. On the other hand the sum of $\delta^\theta_R$ and $\delta^\theta_C$, $\delta^\theta_R+\delta^\theta_C$, becomes $\theta$-independent and zero in the negative energy region due to the cancelation of each other. Therefore, the total phase shift, $\delta^N(E)=N_b\pi+\delta^\theta_R+\delta^\theta_C$, is independent of the scaling angle $\theta$ and finite only in the positive energy ($E>0$) expect for $N_b\pi$. Using this property, we can rewrite the total phase shifts given by integration from $E=0$ as
\begin{eqnarray}
\delta^N(E)=N_b\pi+\pi\int^{E}_{0}\Delta^N_\theta(E)dE. \label{eq3-1-6}
\end{eqnarray}

From the resonance term $\delta^\theta_R$, we can distinguish the 1st and 2nd resonances of the $J^\pi=0^+$ state in Table~\ref{tab1} at the corresponding resonance energies (0.31 MeV and 1.63 MeV), but the structure of higher resonances are not clearly seen. The continuum term $\delta^\theta_C$ has no structure and always negative values indicating a repulsive interaction nature. Its behavior looks like $-a\sqrt{E}$ of a hard sphere scattering. In fact, we can estimate the hard sphere radius $a $ as $4.1$ fm from the behavior of $\delta^\theta_C$ in Fig.~\ref{fig3-1-1}.

In order to see the resonance effect on the phase shift clearly, we calculate the phase shift subtracting the resonance term from $\delta^N(E)$ as
\begin{eqnarray}
\delta^N_{N^\theta_r}(E)&=&\delta^N(E) \nonumber \\
& & -\sum_{r=1}^{N^\theta_r}\int^{E}_{0}dE'\frac{\Gamma_r/2}{(E'-E^{res}_r)^2+\Gamma^2_r/4}.
\end{eqnarray}
In Fig.~\ref{fig3-1-2}, the results of calculation are shown for $N^\theta_r=0,~1,~2,~3,~4$, where $\delta^N(E)=\delta^N_{N^\theta_r=0}(E)$. It is found that the effects of the 1st and 2nd resonances are remarkable, but the 3rd and 4th resonances do not have notable effects. It is shown that every resonance definitely changes the phase shift by $\pi$ at higher energies.

The phase shifts of the $J^{\pi}=1^{-}$ state are also calculated and the results show the similar behavior as the $J^{\pi}=0^{+}$ case for one bound and four resonances shown in Table I.

\subsection{Scattering phase shifts in the $\alpha-\alpha$ and $\alpha-n$ systems}

\begin{figure*}[!tb]
\includegraphics[width=0.80\textwidth,height=0.35\textheight]{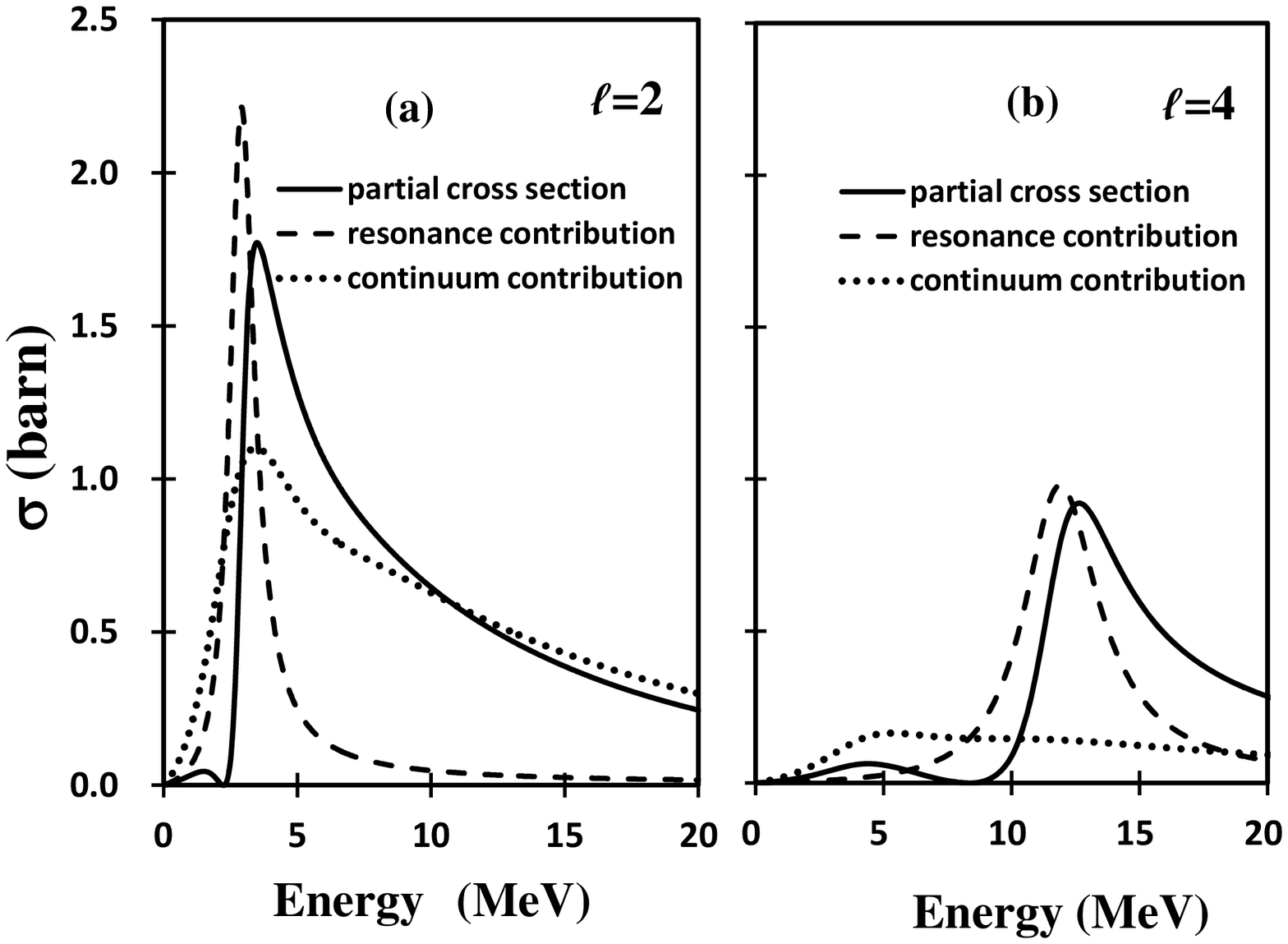}
\caption{The partial cross sections in (a) $\ell=2$ and (b) $4$ waves of the $\alpha-\alpha$ system. The dashed, dotted and solid-lines denote the results of resonance, continuum contributions and partial cross sections, respectively.}
\label{fig3-3-3}
\end{figure*}
\begin{figure}[th]
\includegraphics[width=0.99\columnwidth, height=0.32\textheight]{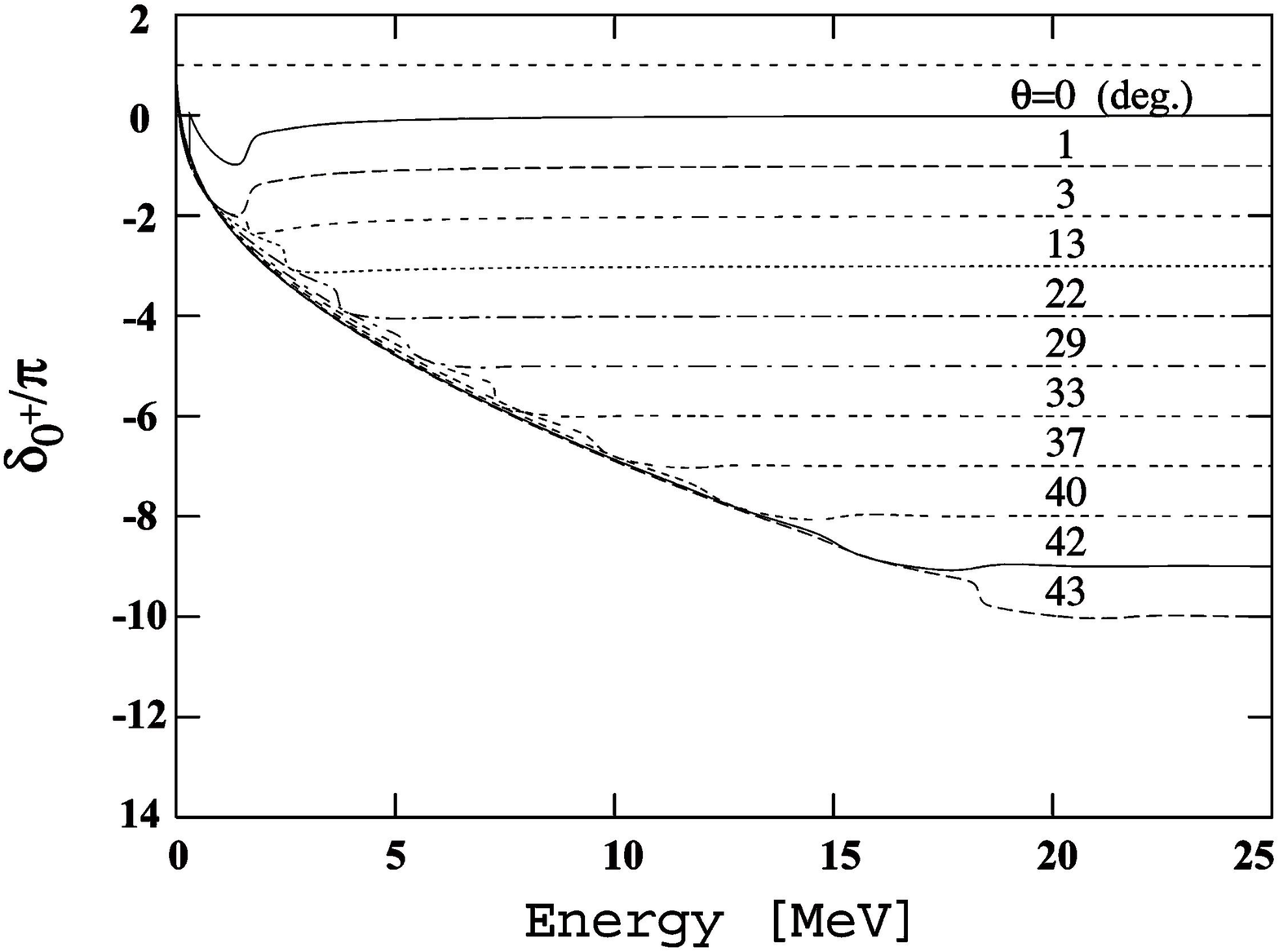}
\caption{The real part of complex phase shift on the '$2\theta$-line' for the $J^{\pi}=0^{+}$ wave at different scaling angles. Energy is given by the absolute value measured from the threshold.}
\label{fig4-1}
\end{figure}
We apply the method to the realistic problems of the $\alpha-\alpha$ and $\alpha-n$ systems comparing with the observed data. The $\alpha-\alpha$ system including a long range Coulomb interaction is described by the following orthogonality condition model (OCM) \cite{Sa69}. Hamiltonians are given as:
\begin{eqnarray}
H&=&-\frac{\hbar^2}{4M}\nabla^2 + V_0\exp{(-\alpha r^2)} + \frac{4e^{2}}{r}\mbox{erf}(\beta r)+\hat{V}_F,\nonumber \\
H_0&=&-\frac{\hbar^2}{4M}\nabla^2 +\frac{4e^{2}}{r},
\label{eq3-2-1}
\end{eqnarray}
where we use $\hbar^2/M=41.471$ MeV$\cdot$fm$^2$, $V_0=-106.09$ MeV, $\alpha=0.75$ fm$^{-1}$ and $\beta=0.22$ fm$^{-2}$. The term $\hat{V}_F$ is a Pauli-potential \cite{Ku84} to project out the Pauli forbidden states from the relative motion in the OCM. The explicit form of $\hat{V}_F$ is given by using the harmonic oscillator wave functions $|n\ell\rangle=u_{n\ell}(r,~b_F)$ as
\begin{equation}
\hat{V}_F=V_F^0\times\left\{
\begin{array}{lll}
|0s\rangle\langle0s|+|1s\rangle\langle1s| &\mbox{for}& \ell=0 \\
|0d\rangle\langle0d| &\mbox{for}& \ell=2 \\
0 &\mbox{for}& \ell\ge 4
\end{array},
\right.\label{eq3-2-2}
\end{equation}
where we use the range parameter $b_F=1.933$ fm and the potential strength $V_F^0=10^6$ MeV. It is noted that the Coulomb potentials between two $\alpha$ particles are expressed by folding potentials with charge distributions of the Gaussian and point in $H$ and $H_0$, respectively.

The eigenvalue problems for these Hamiltonians are solved on the basis of the Gaussian basis function method \cite{Hi03} mentioned above. The resonance solutions for $\ell=0^+,~2^+$ and $4^+$ are obtained as $(E_r^{res},\Gamma_r)=(0.941, 4\times 10^{-5}),~(3.01,1.2),~(11.75,4.4)$ in MeV comparing with the experimental data \cite{Ti04} $(0.0918, (5.57\pm0.25)\times 10^{-6}),~(2.94\pm0.01,1.51\pm0.02),~(11.35\pm0.15,\sim 3.5)$ in MeV, respectively. Using continuum solutions in addition to these resonance ones as shown in lower panels of Fig. \ref{fig3-2-1}, the CLD and the scattering phase shifts are calculated. Here, to see the resonance and continuum phase shifts for positive energies, we redefined them in positive energies;
\begin{eqnarray}
\hat{\delta}_R(E)&=&\int^{E}_{0}dE'\sum_{r=1}^{N^\theta_r}\frac{\Gamma_r/2}{(E'-E^{res}_r)^2+\Gamma^2_r/4},\nonumber \\
\hat{\delta}_C(E)&=&\int^{E}_{0}dE'\left[\sum_{c=1}^{N^\theta_c}\frac{\epsilon^i_c}{(E'-\epsilon^r_c)^2+(\epsilon^{i}_c)^2} \right. \nonumber \\
& & \hspace{1cm} \left.-\sum_{k=1}^{N}\frac{\epsilon^{0i}_k}{(E'-\epsilon^{0r}_k)^2+(\epsilon^{0i}_k)^2}\right].\nonumber \\
\label{eq3-2-3}
\end{eqnarray}

It is noticed that $N_b\pi+\hat{\delta}_R(E)+\hat{\delta}_C(E)=\delta^N(E)$ for $E>0$. The results of $\hat{\delta}_R(E)$ and $\hat{\delta}_C(E)$ are shown in upper panels of Fig.~\ref{fig3-2-1}.

From lower panels of Fig.~\ref{fig3-2-1}, it is seen that one resonant pole in every partial wave of $\ell=0,~2$ and $4$ is obtained. There are no other resonant poles which can make any structure in the resonance phase shifts as seen in upper panels of Fig.~\ref{fig3-2-1}. In the case of $\ell=0$, the resonance term shows a sharp resonance behavior because of the very small resonance width, and the continuum term shows a rather strong repulsive behavior. The $\ell=2$ case shows that the resonance behavior is weakened by the large continuum contribution repulsively. On the other hand, in the $\ell=4$ case, the resonance behavior remains due to small contribution of the continuum term which shows a repulsive nature like other partial waves. This repulsive nature of the continuum terms for $\ell=0$ and $2$ can be understood in association with presence of the Pauli-forbidden states. Because the forbidden states act as bound states to be orthogonal to the scattering solutions, they cause a repulsive nature of the continuum phase shifts.

The $\alpha-n$ system is described by using the Hamiltonians
\begin{eqnarray}
H&=&-\frac{5\hbar^2}{8M}\nabla^2 + V_{\alpha-n}(r) + \hat{V}_F,\nonumber \\
H_0&=&-\frac{5\hbar^2}{8M}\nabla^2,
\label{eq3-2-4}
\end{eqnarray}
where for the $\alpha-n$ potential we use the so-called microscopic KKNN potential \cite{Ka79} given by
\begin{eqnarray}
V_{\alpha-n}(r)&=&\sum_{i=1}^2V_i^C\exp{(-\mu_i^Cr^2)} \nonumber \\
& & \hspace{0.2 cm} +(-1)^{\ell}\sum_{i=1}^3V_{\ell i}^C\exp{(-\mu_{\ell i}^Cr^2)} \nonumber \\
& & \hspace{0.5 cm} +(\mbox{\boldmath$\ell$} \cdot \mbox{\boldmath$s$})[V^{\ell s}\exp{(-\mu^{\ell s}r^2)} \nonumber \nonumber \\
& & \hspace{0.8 cm} +\{1+0.3(-1)^{\ell-1}\}\sum_{i=1}^2V^{\ell s}_{\ell i}\exp{(-\mu^{\ell s}_{\ell i}r^2)}]. \nonumber \\
\label{eq3-2-5}
\end{eqnarray}
The parameters are given in Table~\ref{Table3-2-1}. The Pauli-potential is defined with the harmonic oscillator wave function as
\begin{eqnarray}
\hat{V}_F=V_F^0|0s\rangle\langle0s|,
\label{eq3-2-6}
\end{eqnarray}
where the oscillator size parameter is taken to be $b_F=1.565$ fm and $V_F^0=10^6$ MeV.

\begin{table}[ht]
 \begin{center}
  \caption{Parameters of the $\alpha-n$ KKNN potential \cite{Ka79}.}\label{Table3-2-1}
  \begin{tabular}{l|c|c|c|c}
    \hline\hline
       & $V^C$ (\mbox{MeV})   & $\mu^C$ (\mbox{fm$^{-2}$})   & $V_{\ell}^C$ (\mbox{MeV})   & $\mu^C_{\ell}$ (\mbox{fm$^{-2}$})   \\
    \hline
             & -96.3   & 0.36   &  34.0   & 0.20   \\
  central    &  77.0   & 0.90   & -85.0   & 0.53   \\
             &         &        &  51.0   & 2.50   \\
    \hline
       &  $V^{\ell s}$ (\mbox{MeV})   & $\mu^{\ell s}$ (\mbox{fm$^{-2}$})   & $V_{\ell}^{\ell s}$ (\mbox{MeV})   & $\mu^{\ell s}_{\ell}$ (\mbox{fm$^{-2}$})   \\
    \hline
 spin-orbit  & -16.8 &  0.52  & -20.0   &  0.396  \\
             &         &        &  20.0   &  2.200  \\
    \hline
  \end{tabular}
   \end{center}
\end{table}

Using the Gaussian basis functions, we solve the complex scaled eigenvalue problems for the Hamiltonians of Eq.~(\ref{eq3-2-4}) with $\theta=20^\circ$ and $N=20$ as well. The results for the $p_{3/2}$ and $p_{1/2}$ waves are presented in Fig.~\ref{fig3-2-2}. One resonant pole of the $\alpha-n$ system is obtained: $(E_r^{res},\Gamma_r)=(0.74,~0.59)$ MeV for $p_{3/2}$ and $(2.10,~5.82)$ MeV for $p_{1/2}$, which are compared with the experimental data $(E_r^{res},\Gamma_r)=(0.798,~0.648)$ MeV for $p_{3/2}$ and $(1.27,~5.57)$ MeV for $p_{1/2}$ \cite{Ti02}. Using these results and Eq.~(\ref{eq3-2-3}), we calculate the resonance, continuum and total phase shifts, which are shown in upper panels of Fig.~\ref{fig3-2-2} together with experimental data. We can see a good agreement between theoretical and experimental results. The resonance phase shift of $p_{3/2}$ increases rapidly due to the small decay width. Although $p_{1/2}$ has a larger width, the phase shift of $p_{1/2}$ shows a clear resonance behavior beyond $\pi/2$. The continuum phase shifts of both states are very similar. This trend seems due to the same $p$ wave scattering and a small effect of the $\ell\cdot s$ force to the background states.

The property of the scattering phase shifts is determined from a sum of resonance and continuum terms. Therefore, the observed resonances depend on not only resonant states as poles but also the contribution from the continuum state. The $\alpha-\alpha$ and $\alpha-n$ systems show to keep resonance behavior in spite of existence of continuum contributions. On the other hand, the resonant poles higher than the 3rd one in the Gyarmati potential can not be distinguished as resonances in the phase shifts. They are absorbed in the continuum states.

\subsection{Scattering cross sections of the $\alpha-n$ and $\alpha-\alpha$ systems}
The partial cross sections $\sigma_{\ell} (E)$ are given in Eq.~(\ref{eq2-2-9}), and the total cross section is expressed as
\begin{equation}
\sigma(E)=\sum_{\ell}^{\infty}\sigma_{\ell}(E).
\label{eq3-3-1}
\end{equation}
In Fig.~\ref{fig3-3-1}, we show the comparison of our computed total cross sections with experiments of the $\alpha-n$ system. The experimental data are taken from Ref.~\cite{Au62}. It is found that the calculated total cross sections are in good agreement with the experimental data in a wide energy region. A very sharp peak is observed at the low energy around 2 MeV and has a long tail distribution in higher energies. The low energy cross section dominantly comes from $\ell=1$ partial waves as seen below. The $s$-wave has no resonance due to no barrier. A virtual state corresponding to a pole on the negative imaginary momentum axis can appear in $s$-waves when the interaction between $\alpha$ and neutron has an appropriate strength. However, the KKNN potential is not so strong as to produce such a virtual state \cite{Ma00} and has a repulsive nature for $s$-waves.

Figure~\ref{fig3-3-2} shows the partial cross sections of $p_{3/2}$ and $p_{1/2}$ waves which are decomposed into contributions of resonance and continuum terms, respectively. The partial cross sections are calculated using Eq.~(\ref{eq2-2-9}). Resonance and continuum cross sections are calculated using resonance and continuum phase shifts of $\hat\delta_R(E)$ and $\hat\delta_C(E)$, respectively, in Eq.~(\ref{eq3-2-3}) and are presented in Fig.~\ref{fig3-2-2}. We see that the peak of the total cross section corresponds to the sharp resonance peak in $p_{3/2}$ and the resonance cross section of $p_{1/2}$ gives rather broad distribution. The continuum cross sections show the similar behavior in the $p_{1/2}$ and $p_{3/2}$ waves, which are presented by doted-lines in Fig.~\ref{fig3-3-2}.

In Fig.~\ref{fig3-3-3}, partial cross sections and their decomposition into resonance and continuum terms are shown for $\ell=2$ and $4$ waves of the $\alpha-\alpha$ system. The partial cross section for $\ell=0$ is too sharp as like the delta-function because the small decay width of the $\ell=0$ resonance. Then, we skipped the $\ell=0$ partial cross section. For $\ell=2$ and $4$, resonance cross sections have the shapes like the Breit-Wigner form. The continuum contribution of $\ell=2$ is rather large, while that is small in $\ell=4$. The partial cross section of $\ell=4$ is not so different from the resonance cross section as compared to the $\ell=2$ case. It is interesting that the peak energies of the partial cross section fairly shift from the position of the resonance energies.

\section{Discussion and summary}
The advantage of the CSM is to decompose the unbound states into resonant and non-resonant continuum states by rotating the branch cut on the complex energy plane with a parameter $\theta$. As a result, we can separate a physical quantity into two parts associated with resonant and non-resonant continuum states.
To investigate directly the properties of the continuum states on the '$2\theta$-line' in the CSM, the phase shifts on the '$2\theta$-line' were calculated for the Gyarmati potential \cite{My97}. The phase shifts on the '$2\theta$-line' are obtained as complex values and the real parts are presented for various $\theta$ values in Fig.~\ref{fig4-1}. For $\theta=0$, the calculated phase shift is very similar to the $\delta^N(E)$ in Figs.~\ref{fig3-1-1} and \ref{fig3-1-2}. In Fig.~\ref{fig4-1}, the $\theta$ values are taken for resonant poles so as to be separated one by one from the continuum states. The phase shifts show a jump of $\pi$ at the asymptotic energies due to existence of a resonance between neighboring $\theta$ values. This behavior is completely the same as $\delta^N_{N^\theta_r}(E)$ shown in Fig.~\ref{fig3-1-2}. From this result, we can understand the physical meaning of the '$2\theta$-line' and the solutions obtained along this line in the CSM. In a practical application, we can analyse the scattering quantities decomposing them into sharp resonance term and background term like the relation $\delta^N(E)=\hat{\delta}^\theta_R+\hat{\delta}^\theta_C$ discussed above.

The decomposition of a cross section is not so simple, because it is not equal to a direct sum of resonance and continuum cross sections due to their interference. Since the study by Fano \cite{Fa61}, many discussions have been done so far. Inserting $\hat{\delta}^\theta_R+\hat{\delta}^\theta_C$ into Eq.~(\ref{eq2-2-9}), we have the so-called Fano formula \cite{Sh12}
\begin{eqnarray}
E\sigma_\ell(E)&\propto& \sin^2{(\hat{\delta}^\theta_R+\hat{\delta}^\theta_C)} \nonumber \\
&=&\frac{(q+\epsilon )^2}{(1+q^2)(1+\epsilon ^2)},\label{eq4-1}
\end{eqnarray}
where $\epsilon=-\cot{\hat{\delta}^\theta_R}$ and $q=-\cot{\hat{\delta}^\theta_C}$. In many discussions, $\delta_r$ and $\delta_c$ defined in Eq.~(\ref{eq2-2-7}) have been used instead of $\hat{\delta}_R$ and $\hat{\delta}_C$. When $E_r^{res}>\Gamma_r/2$, they almost coincide with each other. Using the Fano formula, we can understand how the shape of the cross section deviates from the Breit-Wigner form.

In this work we discussed scattering phase shifts and cross sections in the framework of the CSM to investigate the unbound states of two-cluster systems. Using complex energy eigenvalues of resonant and non-resonant continuum states, the analytic form of the phase shifts is derived in a form of sum of resonance and continuum terms in addition to the constant term coming from bound states. This decomposition of phase shifts are very useful to see the resonance contributions in the observed phase shifts and cross sections.

The framework was applied to a simple schematic potential, the so-called Gyarmati potential \cite{Gy90}, producing many resonances for $J^{\pi}=0^{+}$ and $1^{-}$ states. The results indicate that resonances embedded in continuum energies are exposed by using the CSM, and their contributions to the phase shift are made distinctive. This means that the present approach is very promising to understand the role of resonances and their structures in the scattering observables. It is confirmed that the present results of calculation of the phase shift and extracting the resonance terms show a good correspondence with the previous calculations in Ref.~\cite{My97}.

Applying the present framework to the $\alpha-\alpha$ and $\alpha-n$ systems, we obtained the good reproduction of the observed phase shifts and cross sections. The decomposition into resonance an continuum terms makes clear that resonance contributions are dominant but continuum terms and their interference are not negligible. To understand the behavior of observed phase shifts and the shape of the cross sections, both resonance and continuum terms are necessary to be taken into account. If the continuum term is zero, the cross section exhibits a typical Breit-Wigner form. As was discussed by Fano \cite{Fa61}, deviation from the Breit-Wigner form can be investigated by calculating the interference between resonance and continuum terms.

The present method of analyzing the phase shifts and cross sections is also useful in nuclear data evaluations. To study a wider range of nuclear data, it is desirable to develop the method further to include multi-channel systems and to treat many-body systems.\\
\smallskip

\section*{Acknowledgements}

We thank the support by "R$\&$D Platform Formation of Nuclear Reaction Data in Asian Countries (2010-2013)", JSPS AA Science Platform Program. One of the authors, M.O. thanks for hospitality at the Nuclear Reaction Data Centre (JCPRG) and  Theoretical Nuclear Physics Laboratory of Hokkaido University. \\


\begin{thebibliography}{30}
\bibitem{Ho12} H. Horiuchi, K. Ikeda and K. Kat\={o}, Prog. Theor. Phys. Suppl. {\bf 192}, 1 (2012).
\bibitem{Ag71} J. Aguilar and J.M. Combes, Commun. Math. Phys. {\bf 22}, 269 (1971); E. Balslev and J. M. Combes, Commun. Math. Phys. {\bf 22}, 280 (1971).
\bibitem{Ho83} Y. K. Ho, Phys. Rep. {\bf 99}, 1 (1983).
\bibitem{Mo98} N. Moiseyev, Phys. Rep. {\bf 302}, 211 (1998).
\bibitem{Ao06} S. Aoyama, T. Myo, K. Kat\={o} and K. Ikeda, Prog. Theor. Phys. {\bf 116}, 1 (2006).
\bibitem{Ho97} M. Homma, T. Myo, and K. Kat\={o}, Prog. Theor. Phys. {\bf 97}, 561 (1997).
\bibitem{My11} T. Myo, R. Ando and K. Kat\=o, Phys. Letters B {\bf 691}, 150 (2010).
\bibitem{My12} T. Myo, Y. Kikuchi and K. Kat\=o, Phys. Rev. C {\bf 85}, 034338 (2012).
\bibitem{My98} T. Myo, A. Ohnishi and K. Kat\={o}, Prog. Theor. Phys. {\bf 99}, 801 (1998).
\bibitem{My07} T. Myo, K. Kat\=o, H. Toki, and K. Ikeda, Phys. Rev. C {\bf 76}, 024305 (2007).
\bibitem{My01} T. Myo, K. Kat\=o, S. Aoyama and K. Ikeda, Phys. Rev. C {\bf 63}, 054313 (2001).
\bibitem{Su05} R. Suzuki, T. Myo, and K.Kat\={o}, Prog. Theor. Phys. {\bf 113}, 1273 (2005).
\bibitem{Sh92} S. Shlomo, Nucl. Phys. {\bf A539}, 17 (1992).
\bibitem{Le69} R. D. Levine, {\it Quantum Mechanics of Molecular Rate Processes}, 101 (Clarendon Press, Oxford, 1969).
\bibitem{Fa61} U. Fano, Phys. Rev. {\bf 124}, 1866 (1961).
\bibitem{Gy90} A. Cs\'{o}t\'{o}, B. Gyarmati, A. T. Kruppa, K. F. P\'{a}l  and N. Moiseyev, Phys. Rev. A {\bf 41}, 3469 (1990).
\bibitem{Gi03} B. G. Giraud and K. Kat\=o, Ann. of Phys. {\bf 308}, 115 (2003).
\bibitem{Gi04} B. G. Giraud, K. Kat\=o and A. Ohnishi, J. of Phys. A {\bf 37}, 11575 (2004).
\bibitem{My97} T. Myo and K. Kat\={o}, Prog. Theor. Phys. {\bf 98}, 1275 (1997).
\bibitem{Hi03} E. Hiyama, Y. Kino and M. Kamimura, Prog. Part. Nucl. Phys. {\bf 51}, 223 (2003).
\bibitem{Sa69} S. Saito, Prog. Theor. Phys. {\bf 40}, 893 (1968); {\bf 41}, 705 (1969); Prog. Theor. Phys. Suppl. {\bf 62}, 11 (1977).
\bibitem{Ku84} V. I. Kukulin, V. M. Krasnopol'sky, V. T. Voronchev and P. B. Sazonov, Nucl. Phys. {\bf A417}, 128 (1984).
\bibitem{Ti04} D. R. Tilley, J. H. Kelley, J. L. Godwin, D. J. Millener, J. Purcell, C. G. Sheu, and H. R. Weller, Nucl. Phys. {\bf A 745}, 155 (2004).
\bibitem{He10} N. P. Heydenburg, G. M. Temmer, Phys. Rev. {\bf 104}, 123 (1956); C. W. Reich, J. L. Russell, G. C. Phillips, Phys. Rev. {\bf 104}, 135 (1956); C. M. Jones, G. C. Phillips, P. D. Miller, Phys. Rev. {\bf 117}, 525 (1960); T. A. Tombrello, L. S. Senhouse, Phys. Rev. {\bf 129}, 2252 (1963).
\bibitem{Ka79} H. Kanada, T. Kaneko, S. Nagata and M. Nomoto, Prog. Theor. Phys. {\bf 61}, 1327 (1979).
\bibitem{Ti02} D. R. Tilley, C. M. Cheves, J. L. Godwin, G. M. Hale, H. M. Hofmann, J. H. Kelley, C. G. Sheu, H. R. Weller, Nucl. Phys. {\bf A708}, 3 (2002).
\bibitem{Ho66} B. Hoop Jr and H. H. Barschall, Nucl. Phys. {\bf 83}, 65 (1966); Th. Stammbach and R. L. Walter, Nucl. Phys. {\bf A180}, 225 (1972).
\bibitem{Au62} S. M. Austin, H. H. Barschall and R. E. Shamu, Phys. Rev. {\bf 126}, 1532 (1962); R. E. Shamu and J. G. Jenkin, Phys. Rev. {\bf 135}, B99 (1964); F. J. Vaughn, W. L. Imhof, R. G. Johnson and M. Walt, Phys. Rev. {\bf 118}, 683 (1960); Los Alamos Physics and Gryogenics Groups, Nucl. Phys. {\bf 12}, 291 (1959).
\bibitem{Ma00} H. Masui, S. Aoyama, T. Myo, K. Kat\=o and K. Ikeda, Nucl. Phys. {\bf A 673}, 207 (2000).
\bibitem{Sh12} I. Shimamura, Adv. Quant. Chem. {\bf 63}, 165 (2012).

\end{thebibliography}
\end{document}